\documentclass[twocolumn,showpacs,floatfix,superscriptaddress,prl]{revtex4}
\usepackage{graphicx,amsfonts,amssymb,amsmath, hyperref}
\usepackage[applemac]{inputenc}
\usepackage{ifpdf}

\newif\ifhyper

\hypertrue
\ifhyper
\hypersetup{
  citecolor = {green},
  colorlinks = {true}, % false
  urlcolor = {blue} % magenta
}
\fi

\newlength{\ldag}
\newcommand{\be}{\begin{equation}}
\newcommand{\ee}{\end{equation}}         
\newcommand{\ba}{\begin{array}}
\newcommand{\ea}{\end{array}}

\begin{document}

\title{First order phase transitions in  polymerized  phantom membranes}

\author{K. Essafi} \email{karim.essafi@oist.jp}
\affiliation{OIST, 1919-1 Tancha Onna-son, Okinawa 904-0495 Japan}
\affiliation{Sorbonne Universités, UPMC Univ  Paris 06, UMR 7600, LPTMC, F-75005, Paris, France}

\author{J.-P.  Kownacki}\email{kownacki@u-cergy.fr}
\affiliation{LPTM, CNRS UMR 8089 - Universit\'e de Cergy-Pontoise,   2 avenue Adolphe Chauvin, 95302 Cergy-Pontoise  Cedex, France}

\author{D. Mouhanna} \email{mouhanna@lptmc.jussieu.fr}
\affiliation{Sorbonne Universités, UPMC Univ  Paris 06, UMR 7600, LPTMC, F-75005, Paris, France}

\begin{abstract}

The crumpled-to-flat phase transition that occurs  in   $D$-dimensional polymerized phantom membranes embedded in a $d$-dimensional space  is investigated nonperturbatively  using a  field expansion  up to order eight in powers of  the order parameter.   We   get  the critical dimension $d_{\rm cr}(D)$ that separates a second order region   from a   first order one everywhere between $D=4$ and $D=2$. Our  approach  strongly suggests  that the phase transitions   that take place in physical membranes  are  of  first order in   agreement  with most  recent  numerical simulations.

\end{abstract}


\maketitle

%\tableofcontents

Fluctuating or random  surfaces  are  a recurrent  concept  in  physics \cite{proceedings89,bowick01}.  They  occur  in  soft matter  physics or   in biology as   assemblies  of  amphiphilic molecules that  can form  plane  or  closed  structures  (vesicles)   according to  the chemical composition  of  the membrane itself and its  surroundings.    Random  surfaces  also  appear  in high-energy physics, especially  in string theory,  as the world-sheet swept out  by a string  during its  spacetime evolution.   More recently  membranes have received  a renewed interest in condensed matter physics where it has been realized that, from the point of view of  their mechanical properties, novel materials,  like graphene \cite{geim07},  identify  with polymerized membranes, providing  the first and unique example of genuinely  two-dimensional membrane \cite{meyer07,katsnelson13}.   The coexistence of  two-dimensional  geometry and thermal  fluctuations   is   at  the origin of  a  variety of   behaviours depending on  the nature of  the internal structure of the  membrane.   {\it Fluid} membranes  are made  of molecules that  freely diffuse and  re-arrange rapidly when a shear or a stress  is performed.    This implies that, in absence of an  external tension,   the dominant  energy is the  bending energy.  It  has  been shown that, in this case,  the height fluctuations are sufficiently strong to prevent  the appearance of long-range order; fluid membranes are thus always crumpled \cite{helfrich85,peliti85}.   {\it Polymerized} -- or tethered -- membranes  display a  drastically different behaviour.  Indeed   the existence of an underlying  network of linked molecules   induces   elastic (shearing  and stretching) energy contributions  that lead to a coupling between   height  and transverse -- phonons -- modes.  It results from this situation   a {\it frustration}   of the height fluctuations \cite{nelson04} that are strongly  reduced at low temperatures giving rise  to  the appearance of a  flat  phase  with long-range order between the normals  \cite{nelson87,david88}.     The existence of  a low-temperature phase accompanied with  spontaneous symmetry breaking of rotational invariance is {\it a priori}  in contradiction with the Mermin-Wagner theorem.  However  it appears that  the effective phonon-mediated interaction between the height fields  (more precisely between the Gaussian curvatures) is  of long-range kind, allowing to evade  the conditions of application of the Mermin-Wagner theorem \cite{nelson87}.  Correlatively the   low-temperature phase of membranes  is  characterized by non trivial scaling behaviour in the infrared \cite{aronovitz88,guitter89,aronovitz89}:
 \begin{equation}
\left\{
\begin{array}{ll}
G_{hh}({\bf q})\underset{q\to 0}{\sim}  q^{-(4-\eta)}\\
\\
G_{uu}({\bf q})\underset{q\to 0} {\sim}  q^{-(6-D-2 \eta)}
\nonumber
\end{array}
\right.
\end{equation}
where $G_{hh}({\bf q})$ and $G_{uu}({\bf q})$ are the correlation functions  of  the out-of-plane and in-plane excitations  respectively.  The exponent  $\eta$ characterizes    stable membranes  including,   for   instance,    sheets  of graphene.  The determination of this exponent   has been the object of an intense activity.  Early  perturbative approaches  performed  below   the upper critical dimension $D=4$ at first order in $\epsilon=4-D$  have led -- with $\epsilon=2$ --  to the questionable  value $\eta\simeq 0.96$ \cite{aronovitz88}. On the other hand    large-$d$  expansion  performed at lowest order   and  extrapolated to the  finite value    $d=3$  has produced  another doubtful  value:     $\eta=2/3$  \cite{david88,aronovitz89}.  More  sophisticated computations performed in the aftermath have revealed   a remarkable stability of the results with the  order of the approximations. For instance self-consistent screening approximations (SCSA)  at  leading and next-to-leading order  have  led   to   close  values:   $\eta\simeq 0.821$ \cite{ledoussal92,gazit09,zakharchenko10,roldan11}  and $\eta=0.789$ \cite{gazit09}  respectively.    Nonperturbative renormalization group (NPRG) approaches  have also    been used, first within a double field and field-derivative expansion leading to the value   $\eta\simeq 0.849$ \cite{kownacki09}. This  computation  has  been  extended    by   including  the  momentum dependence  of the vertices that were considered in Ref.\cite{kownacki09}. Very surprisingly   an almost unchanged value of  $\eta$ has been obtained:  $\eta\simeq 0.85$  
\cite{braghin10,hasselmann11}.  This    exponent  has  also  been  computed by   means of  Monte Carlo  simulations  of membranes  leading to the values $\eta=0.81(3)$ \cite{zhang93} and  $\eta= 0.750(5)$ \cite{bowick96},   and recently using  a  Fourier Monte Carlo simulation the value $\eta=0.795(10)$ \cite{troster13}.  Finally  Monte Carlo simulations of   graphene   using   a realistic potential   have concluded in favor of the  value $\eta\simeq 0.85$ \cite{los09}  in agreement with  the SCSA and even better with NPRG approaches.   This agreement between the  field  theoretical approaches and the Monte Carlo computations  of  membranes or materials like  graphene  has proven    the relevance  of  the continuum models to describe the long-distance behaviour of these systems  in the flat phase. It has also  shown   the adequacy of the approximations used  within  these field theoretical approaches -- see below    for  considerations about  the NPRG computations.

On the opposite the question of the nature of the crumpled-to-flat transition is still  largely  open.   In this context several perturbative  approaches have also   been  used  to investigate the critical physics of $D$-dimensional polymerized membranes embedded in a $d$-dimensional space.  A weak-coupling perturbative approach performed in the vicinity of  the upper critical dimension $D=4$ \cite{paczuski88} has  led to predict a  second order  phase transition for $d>d_{\rm cr}$ and  first order phase transitions for $d<d_{\rm cr}$ with $d_{\rm cr}\simeq 219$ just below  $D=4$.      However, again, these results are  not directly relevant for  two-dimensional membranes.  Alternatively $1/d$ expansion  \cite{david88,aronovitz89} or SCSA  \cite{ledoussal92}    have been  employed  to investigate this  transition. However a main  problem with these approaches  is their inability to identify   the existence -- and thus to get the value -- of the critical dimension $d_{\rm cr}(D)$ separating the  first from the second order region.  Finally the crumpled-to-flat transition has been investigated by means of Monte Carlo   and cluster variation method (CVM) computations for membranes or related models.  While early computations favored a second order phase transition (see \cite{kantor04,gompper04,bowick01} for reviews) more  recent computations    exhibit  clear  first order behaviours \cite{kownacki02,koibuchi04,koibuchi05,koibuchi14}, see also \cite{koibuchi08,nishiyama10,cirillo12}.

Recently  NPRG approaches have been used to investigate both the flat phase and  the nature of the crumpled-to-flat  transition \cite{kownacki09,braghin10,hasselmann11}.   These approaches   are based  on the concept of   running effective action  \cite{wetterich93c}  (see \cite{bagnuls01,berges02,delamotte03,pawlowski07,rosten12}  for reviews),  $\Gamma_k[\boldsymbol \phi]$,   a  functional of   the  order parameter  field 
$\boldsymbol \phi$,  that describes the effective physics  at a  coarse-grained 
 scale $k$. Technically   the index  $k$  stands for  a running scale that separates the high-momentum modes,  with $q>k$,  from  the low-momentum ones, with $q<k$ and     $\Gamma_k[\boldsymbol \phi]$  represents   a coarse grained free energy where only fluctuations with momenta $q>k$  have been integrated out.  The   running  of $k$ towards   $k=0$ thus corresponds to  gradually integrating over all fluctuations. The $k$-dependence, RG flow,  of $\Gamma_k$ is  provided  by an exact -- albeit one-loop -- evolution equation  now known as the Wetterich equation \cite{wetterich93c}.  Solving this equation allows to get   the RG flows and thus  the critical properties.   However  although exact the Wetterich equation cannot be solved exactly and approximations must be performed  in order to make  the computations tractable.   Several  kinds of approximations are possible and some of them have already  been used in the context of membranes:  {\it i)} a double expansion of the effective action  in powers of the field and derivatives of the field   \cite{kownacki09} {\it ii)}  a field expansion  of the effective action where the full momentum dependence of the vertices  is kept  \cite{braghin10,hasselmann11}.  On the  basis  of these approaches,  that have been performed  up to  order  four in powers of the field, a second order phase transition has been predicted for polymerized membranes,   allowing nevertheless the   possibility  of   first order transitions.  A crucial point within the NPRG context is  to guess the  suitable approximation  of the effective action  $\Gamma_k[\boldsymbol \phi]$ and   to  check the convergence of the results when enriching its  content.     We address here the question of the relevance of  the  high-order vertices that have been  discarded   in  both  \cite{kownacki09}  and in  \cite{braghin10,hasselmann11}.   To do this we consider  an expansion up to order eight  in powers of the field,  keeping only their local -- zero momentum -- part,   extending  thus  the computation   performed in \cite{kownacki09}.    Our motivation is  threefold: {\it i)} to probe  the importance of  terms  that, from the point of view of standard power-counting,  are  the most  relevant -- and in particular more relevant  than the  non-local ones considered in \cite{braghin10,hasselmann11}  {\it ii)}  to optimize our results using different kinds of cut-off families {\it iii)}  to test the  convergence of our results  with respect to  the field content.    We  show  that  the terms considered here drastically affect the  critical behaviour  since  they  very likely  turn  the second order  phase transition  into  first order  ones.

We consider a   $D$-dimensional membrane embedded in a  $d$-dimensional Euclidean space.   The location of a point on the membrane
is realized  by the use of   $D$-dimensional {\it internal}   coordinates ${\bf x}\equiv x_{\mu}$,  $\mu=1\dots  D$ while a configuration of the membranes in the Euclidean space is realized through the embedding:  ${\bf x}\to {\bf R} ({\bf x})$ with ${\bf R}\equiv (R^i), i=1\dots  d$.  The energy of a  polymerized membrane  is made  of  a bending  energy part   and a stretching energy part.   In the spirit of a   Landau-Ginzburg-Wilson  approach this energy can be expressed as an expansion in powers of the microscopic field  $\bf R$ and  its derivatives.  Here translational invariance in the embedding space  and rotational invariance --  in the embedding  as well as in the internal   space --  of  the energy  lead to  an expansion in terms of the tangent vectors $\partial_{\mu} {\bf R}$,  $\mu=1\dots  D$.  The energy is given by: 
\begin{equation}
\begin{array}{ll}
\displaystyle  H \left[{\bf R}\right]=\displaystyle \int  d^ D x\  \displaystyle{\kappa\over 2} \left(\partial_{\mu}\partial_{\mu}\bf{R}\right)^2  
+ {t\over 2} \left(\partial_{\mu}{\bf R}\right)^2    + \\
\\
\hspace{1.5cm} +\ \    u\left(\partial_{\mu}{\bf R}. \partial_{\nu}{\bf R}\right)^2 +  v\left(\partial_{\mu}{\bf R}. \partial_{\mu}{\bf R}\right)^2 + \dots 
\label{hamilton}
\end{array} 
\end{equation}
with $\partial_{\mu}={\partial/ \partial x_{\mu}}$,  $\mu=1\dots  D$. In (\ref{hamilton}) the first term represents  the bending energy  with   $\kappa$ the rigidity constant, the second one a tension while the other terms correspond  to  stretching energy with  $u$ and $v$ being the Lamé coefficients.  Finally note that the temperature $T$ has been absorbed in the bare coupling constants. Let us consider a mean-field approach of the crumpled-to-flat phase transition, with the assumptions $u>0$ and $u+v D>0$. 
For $t>0$  the minimum  of  (\ref{hamilton}) is given by  a crumpled phase with a vanishing  average value of the order parameter:  $\partial_{\mu}{\bf r}\,  \hat=\,  \langle \partial_{\mu}{\bf R}\rangle=0$  expressing  the absence of long-range orientational order.  For  $t<0$  the minimum of   (\ref{hamilton}) is given by  non-vanishing values of: 
\begin{equation}
\partial_{\mu}{\bf r} =\langle \partial_{\mu}{\bf R}\rangle=\zeta  {\bf e_{\mu}}
\label{flat}
\end{equation}
$\mu=1\dots  D$ where $\zeta=\sqrt{-t/4(u+v D)}$ and where  the ${\bf e_{\mu}}$ form   an orthonormal set of  $D$  vectors -- $ {\bf e_{\mu}}. {\bf e_{\nu}}=\delta_{\mu\nu}$ -- that span  the  flat  phase.  This occurrence of long-range orientational order with non-vanishing values  of the  mean  tangent vectors $\partial_{\mu}{\bf r}$  at low-temperature is  analogue to the occurrence of ferromagnetism in spin systems,  the tangent vectors playing  the role of  order parameters with magnitude given by $\zeta$.  Thus at $t=0$, in the mean-field approximation,  there is a phase transition between a high-temperature, {\it crumpled},  phase and a low-temperature, {\it flat},  phase displaying long-range orientational order.

From now on we employ  the effective action, obtained through a Legendre transform of the free energy \cite{berges02},  in terms of which the Wetterich equation is written. 
We consider a  derivative expansion of the effective action: 
\begin{equation}
\Gamma_k[{\bf r}] = \displaystyle \int d^Dx\  \ {Z_k\over 2} (\partial_{\mu}\partial_{\mu} {\bf r})^2 + U_k[\partial_{\mu}  {\bf r}] + O(\partial_{\mu}^{\,6})\ . 
\nonumber
\end{equation}
In this expression $Z_k$ is a field-independent  field renormalization,  $U_k[\partial_{\mu}  {\bf r}]$ is the potential part of the effective action that we  expand in powers of the field  $\partial_{\mu}  {\bf r} $ around the flat phase configuration (\ref{flat}):
\begin{equation}
U_k[\partial_{\mu}  {\bf r}] = \displaystyle  \sum_{n_1,\dots,n_D\ge 0} a_{n_1,\dots,n_D}({\hbox{Tr}} [g])^{n_1}\dots ({\hbox{Tr}}[g^D])^{n_D}
\label{potential}
\end{equation}
where $g$ stands for the -- metric --  tensor  with elements:  $g_{\mu\nu}=\partial_{\mu}{\bf r}.\partial_{\nu}{\bf r}-\zeta^2 \delta_{\mu\nu}$. We have considered the expansion (\ref{potential}) up to order eight in the field, which means  to follow the flow of 10  coupling constants.  Note that the engineering  dimension of the  order parameter   $\partial_{\mu}  {\bf r}$  is  given by:  $[\partial_{\mu}  {\bf r}]=(D-2)/2$. The   dimension   of a generic  coupling constant entering  in (\ref{potential}) is  thus given by $[a_{n_1,\dots,n_D}]=D-(D-2)(n_1+2 n_2+ \dots + D n_D)$.  Therefore the  coupling constants  of  (\ref{potential}) are all relevant with dimension $2$ in $D=2$. This is in contrast with the terms considered   in \cite{hasselmann11} which  involves at least two additional  derivatives and are  thus  at best marginal.  
One can thus  anticipate    from this  power counting that the terms considered  here  play a  more  significant role than those treated   in  \cite{hasselmann11}.

The RG flow of  $\Gamma_k$ is provided  by  the Wetterich equation \cite{wetterich93c}:
\begin{equation}
{\partial \Gamma_k\over \partial t}={1\over 2} \hbox{Tr} \left\{{\partial R_k\over \partial t} (\Gamma_k^{(2)}+R_k)^{-1}\right\}
 \label{renorm}
\end{equation}
where $t=\ln \displaystyle {k / \Lambda}$,  $\Lambda$ being  some ultraviolet scale. The trace has to be understood as a $D$-dimensional momentum  integral as well as a summation over internal indices.  In Eq.(\ref{renorm}),  $\Gamma_k^{(2)}$  is the  inverse propagator, the second derivative of $\Gamma_k$ with respect to the field $\bf r$, taken in a {\sl generic}, nonvanishing field configuration.  $R_k(q)$ is  a  cut-off function that suppresses  the propagation of modes with momenta $q<k$ and makes that $\Gamma_k$ encodes only modes with momenta $q> k$. It also regulates the ultraviolet behaviour.  For our study we  have used three  cut-off families parametrized by a real number  $\lambda$:  $R_{k,1}^{\lambda}(q)=Z_k\,  \lambda (k^4-q^4)\theta(k^2-q^2)$, $R_{k,2}^{\lambda}(q)=Z_k\, \lambda\,  q^4 /({\hbox{exp}}(q^4/k^4)-1)$ and $R_{k,3}^{\lambda}(q)=Z_k\, \lambda\, {\hbox{exp}}(-q^4/k^4)$. The RG equations  for  the  different coupling constants  entering in (\ref{potential}) are obtained by expressing  them  as  functional derivatives  of  $\Gamma_k[\bf r]$, taking their  $t$-derivatives  and using (\ref{renorm}). We have successively considered the  order  four, six and eight  of the expansion in powers of  $\partial_{\mu}{\bf r}$, with the eight  order   appearing  as  the last order   manageable within  the current   abilities  and a reasonable period of time   \footnote{The $\beta$  functions at  next order(s) appear  to be  impossible to derive  due  to the  extremely rapid growth  of the size of the expressions we need to deal with. }.   The  explicit equations being  too long to be displayed here we only  provide  the results.  

{\it Flat phase.}  Within our formalism the   RG equations in this phase are  obtained by taking the limit  where the dimensionless magnitude  $\bar \zeta$ goes to infinity  (see \cite{kownacki09})  corresponding  to the decoupling of the phonon modes and  to a spectrum of excitations dominated by the height fluctuations.  Our analysis shows that  the RG equations obtained in the flat phase  at order four  in  \cite{kownacki09} are not modified by  {\it any}  higher  power  of  the field (see \cite{essafi14b} for technical details).  Thus the results derived in \cite{kownacki09}, and in particular the value of $\eta\simeq 0.849$,   are correct at {\it all orders}  within an expansion in powers of  $\partial_{\mu} {\bf r}$. This fact  explains  in a great extent the agreement between our  prediction for $\eta$ and that obtained with the Monte Carlo simulation. It remains however to understand also why the derivative terms considered in \cite{hasselmann11}  do almost not modify the  value of $\eta$.

\begin{figure}[htbp]
\vspace{0.2cm}\hspace{-0.5cm}
{\includegraphics[width=8cm,height=7cm]{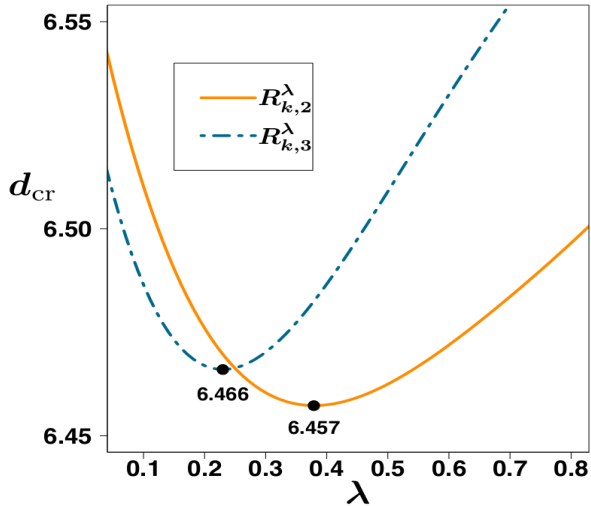}}
\vspace{0cm}
\caption{Optimization of $d_{\rm cr}(D=2)$ as function of the parameter $\lambda$ for the approximation of order six  for the cut-off functions $R_{k,2}$ and   $R_{k,3}$.}
\label{optimal}
\end{figure}

\begin{figure}[htbp]
\vspace{0.1cm}\hspace{-0.5cm}
{\includegraphics[width=7cm,height=7cm]{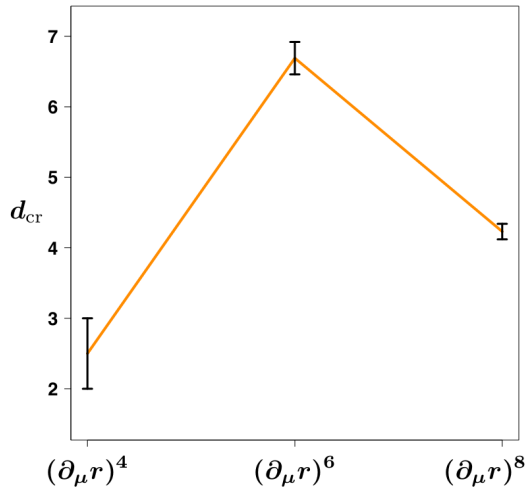}}
\vspace{0cm}
\caption{Critical dimension $d_{\rm cr}(D=2)$ as a function of the order of the expansion.}
\label{critical}
\end{figure}

\begin{figure}[htbp]
\vspace{0cm}\hspace{-0.5cm}
{\includegraphics[width=8cm,height=6.5cm]{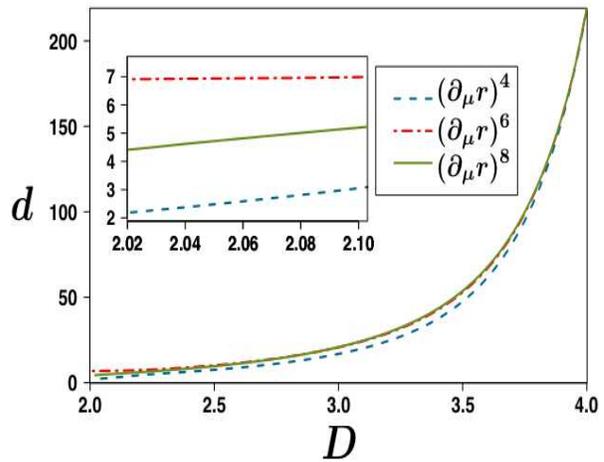}}
\vspace{0cm}
\caption{Curves $d_{\rm cr}(D)$  at different   orders  of the expansion evaluated with the cut-off $R_{k,1}$.}
\label{criticalD}
\end{figure}

{\it Crumpled-to-flat transition.} The situation is very different  for the crumpled-to-flat transition.  At each order, four, six and eight  in powers of  the field we have determined the value of $d_{\rm cr}(D)$.  We have used the three families of cut-off functions   $R_{k,1}$, $R_{k,2}$ and   $R_{k,3}$ and  optimized the result by applying  the principle of minimal sensitivity, see \cite{canet03a,canet03b}. For almost each  family of cut-off we have found  a marked minimum  of $d_{\rm cr}(D)$ as a function of the parameter $\lambda$. This is illustrated at order six  in powers of the field and for $D=2$  in Fig.\ref{optimal}.  Then, at each order,   we have averaged  over the different values obtained from different cut-off families.  The results are  displayed in Fig.\ref{critical},  the  bar  corresponding to the dispersion of values of $d_{\rm cr}(D=2)$.    As can be seen from  Fig.\ref{critical},    $d_{\rm cr}$   displays an oscillating behaviour with  amplitudes  decreasing   with the order of the expansion.  This damped oscillating  behaviour  is typically observed  for critical exponents  or other  physical quantities in various  systems within the field expansion \cite{tissier00,tissier01,tissier01b,canet03a,canet03b}. This  strongly suggests    a convergence  toward a  value  located between   4.5 and 6.5, excluding  in particular the value $d=3$ and thus  a second  order phase transition.  This conclusion is strengthened  by the fact that  the dispersion of results  when varying the cut-off functions  decreases with the order of the expansion.   Finally  Fig.\ref{criticalD} displays the  shape   of the curve $d_{\rm cr}(D)$  everywhere between the upper critical dimension $D=4$ and $D=2$ computed  for the cut-off  $R_{k,1}$.

%%%%%%%%%%%%%%%%%%%%%%%%%%%%%%%%%%%%%%%%%%%%%%%%%%%%%%%%%%%%%%%%
%%%%%%%%%%%%%%%%%%%%%%%%%%%%%%%%%%%%%%%%%%%%%%%%%%%%%%%%%%%%%%%%

 %%%%%%%%%%%%%%%%%%%%%%%%%%%%%%%%%%%%%%%%%%%%%%%%%%%%%%%%%%%%%%%%
%%%%%%%%%%%%%%%%%%%%%%%%%%%%%%%%%%%%%%%%%%%%%%%%%%%%%%%%%%%%%%%%

%%%%%%%%%%%%%%%%%%%%%%%%%%%%%%%%%%%%%%%%%%%%%%%%%%%%%%%%%%%%%%%%
%%%%%%%%%%%%%%%%%%%%%%%%%%%%%%%%%%%%%%%%%%%%%%%%%%%%%%%%%%%%%%%%

{\it Conclusion}. Taking into account orders beyond the power four of the field  within  a field expansion deeply modifies the critical behaviour of  polymerized membranes at the crumpled-to-flat transition, contrary to what happens for the flat phase that displays a strong stability. Our  study  favours the existence of  first order phase transitions for  phantom polymerized membranes in agreement with most recent numerical  computations  \cite{koibuchi14}. Note that  since the  order  eight  appears as  the last manageable order  of the field expansion a definitive conclusion would certainly require  a full treatment of the potential part  of the effective action as well as  managing  (a part of) derivative terms \cite{essafi14b}. However,  the behaviour of the critical dimension $d_{\rm cr}$ with the order of the field expansion,  combined with knowledge acquired from other systems, makes  highly unlikely a change of the nature  of the transition compared with the present results.

%\bibliographystyle{unsrt}
%\bibliography{bibliotheque1}

\end{document}